\documentclass[twocolumn,amsmath,amssymb,prd,reprint,longbibliography,floatfix,8pt]{revtex4-1}
\usepackage{cancel}
\usepackage{enumitem}
\usepackage[utf8x]{inputenc}
\usepackage{graphicx}
\usepackage{dcolumn}
\usepackage{braket}
\usepackage{bm}
\usepackage{hyperref}
\usepackage[switch]{lineno}
\usepackage{float}
\usepackage{subfigure}
\usepackage{tikz}
\usepackage{multirow}
\usepackage{amsmath}
\usepackage{chemformula} 
\usepackage[T1]{fontenc} 
\usepackage{makecell}
\usetikzlibrary{arrows.meta}
\usetikzlibrary{arrows,decorations.pathmorphing,backgrounds,positioning,fit,petri}

\begin{document}

\title{Generating new coordination compounds via multireference simulations, genetic algorithms and machine learning: the case of Co(II) molecular magnets}
\author{Lion Frangoulis}    
\email{Authors contributed equally}
\affiliation{School of Physics, AMBER and CRANN Institute, Trinity College, Dublin 2, Ireland}
\author{Zahra Khatibi}
\email{Authors contributed equally}
\affiliation{School of Physics, AMBER and CRANN Institute, Trinity College, Dublin 2, Ireland}
\author{Lorenzo A. Mariano}
\affiliation{School of Physics, AMBER and CRANN Institute, Trinity College, Dublin 2, Ireland}
\author{Alessandro Lunghi}
\email{lunghia@tcd.ie}
\affiliation{School of Physics, AMBER and CRANN Institute, Trinity College, Dublin 2, Ireland}

\begin{abstract}
    The design of coordination compounds with target properties often requires years of continuous feedback loop between theory, simulations and experiments. In the case of magnetic molecules, this conventional strategy has indeed led to the breakthrough of single-molecule magnets with working temperatures above nitrogen's boiling point, but at significant costs in terms of resources and time. Here, we propose a computational strategy able to accelerate the discovery of new coordination compounds with desired electronic and magnetic properties. Our approach is based on a combination of high-throughput multireference ab initio methods, genetic algorithms and machine learning. While genetic algorithms allow for an intelligent sampling of the vast chemical space available, machine learning reduces the computational cost by pre-screening molecular properties in advance of their accurate and automated multireference ab initio characterization. Importantly, the presented framework is able to generate novel organic ligands and explore chemical motifs beyond those available in pre-existing structural databases. We showcase the power of this approach by automatically generating new Co(II) mononuclear coordination compounds with record magnetic properties in a fraction of the time required by either experiments or brute-force ab initio approaches. 
\end{abstract}

\maketitle

\section*{Introduction}

Coordination compounds of first-row transition metals and lanthanide ions often result in molecules with a large magnetic moment \cite{zabala2021single,sessoli1993magnetic,gomez2015large}. If properly tuned, the magnetic properties of these compounds can lead to a slow reorientation of their magnetic moment, resulting in a molecular counterpart to permanent magnets \cite{metal-radical, rise_of_3d,Lanthanide_overview}. These molecular compounds, generally referred to as single-molecule magnets (SMMs), have been under intense scrutiny for different applications, including nano-magnetism \cite{friedman2010single}, high-density information storage \cite{raju2024optical}, spintronics \cite{bogani2008molecular}, and quantum technologies \cite{atzori2019second}. As for bulk permanent magnets, easy-axis magnetic anisotropy is central to the observation of the slow relaxation of the magnetic moment \cite{Lunghi2022}. At the molecular scale, this property is often measured in terms of the ion's axial zero-field splitting $D$ \cite{neese2007calculation}. For large negative values of this figure of merit, a given direction in space becomes energetically favorable for the molecular magnetic moment, protecting it from thermal fluctuations induced by the coupling of the magnetic moment with the molecular crystal vibrations \cite{lunghi2022computational}. Following this design principle, hundreds of coordination compounds have been synthesized, and the highest working temperature of SMMs has risen from 2 to 80 K since the first report of this phenomenon \cite{guo2018magnetic,yin2023recent}.  

Despite this incredible breakthrough, the road leading to it has taken 30 years of developments in synthetic chemistry, theoretical models and computational methods, begging the question of how much longer it will take for the next significant step forward. This is not a problem of molecular magnetism alone, and enabling accelerated molecular or materials discovery is a pressing need across the full spectrum of technological and biomedical applications. Indeed, most ideal materials often represent a sparse and small subset of a virtually infinite chemical space of candidates, making a serendipitous approach to materials discovery unsuitable for scientific advancements at a sustainable cost and environmental impact \cite{gutierrez2025call,lannelongue2021green}.

In this space, computational methods are increasingly prominent. Ab initio simulations, in particular, have become of central importance by allowing accurate predictions of many molecular properties, but their application to a large number of different molecules and materials requires additional care, lest computational overheads become prohibitive \cite{curtarolo2013high}. Notable approaches to achieve this are high-throughput \cite{gomez2016design, shi2020machine, Compass, chew2025leveraging}, evolutionary \cite{GA_book_Holland, GA_polymer, GA_Self_Assembly, GA_Review1, GA_review2, GA_Overview, GA_in_chemometrics} and machine-learning methods \cite{simonovsky2018graphvae, popova2018deep, gomez2018automatic, jin2020chapter, zeni2023mattergen, jin2024liganddiff,strandgaard2024deep, du2024machine}. These computational schemes, despite significant differences, all follow a similar underlying philosophy: past information about the molecular properties of interest can be exploited to speed up the sampling of new relevant molecules. High-throughput methods achieve this by screening as many compounds as possible for a suitable, easy-to-compute descriptor linked to the properties of interest, machine learning provides a way to predict molecular properties at low computational cost by training a statistical model from existing datasets, and evolutionary methods prescribe an informed guess on what molecules are likely to exhibit the property of interest based on known best candidates. 

These strategies have found widespread application in the design of either organic and biological compounds \cite{GA_Drug1, GA_Drug2, GA_Drug3, GA_opt} or solid-state inorganic crystals \cite{KARIUKI1997189, Abraham2006, Wu_2014, Xtal}. However, due to their inherent complexity, coordination compounds have received limited attention so far, with most exceptions revolving around the structural optimization of catalysts \cite{GA_Cata1, GA_Cata2, GA_TMC, GA_Geo_Opt, GA_Catalysis3, GA_Cata4}. Filling such a methodological gap comes with several challenges, particularly if electronic and magnetic properties are of interest \cite{nandy2021computational}. Different from organic compounds, coordination ones have, on average, larger numbers of atoms and a much broader range of bond motifs, making it hard to represent them with computer algorithms and to access their full chemical space. On top of this, the screening for electronic and magnetic properties adds complexity by requiring high-level multireference ab initio methods to compute them to the desired accuracy \cite{feldt2022ab}. Recently, a study has shown that multireference electronic structure can be used in an automated fashion to perform high-throughput simulations of magnetic molecular properties \cite{Compass}, but two main roadblocks still stand between us and a real automated and efficient workflow that is able to navigate arbitrarily large portions of the chemical space: i) only a small fraction of compounds randomly sampled exhibit relevant properties, making random sampling of molecular candidates extremely inefficient, and ii) the chemical space that can be explored is restricted to the organic ligands present in crystallographic databases, strongly limiting the search for rare outstanding candidates \cite{Compass}. 

In this study, we explore a workflow that brings together the strengths of ab initio methods, genetic algorithms (GAs) and machine learning to tackle these outstanding challenges. We apply this framework to the generation of novel mononuclear coordination compounds of Co(II) with optimal values of magnetic properties. This problem represents an ideal testbed that combines the genuine potential for experimental realization and technological breakthrough with the possibility of comparing results with previous brute-force high-throughput simulations and extensive literature. Results show that this framework is able to generate new coordination compounds of Co(II) with excellent values of $D$ in just a few hundred ab initio calculations and close to the theoretical limit in just about one thousand calculations, paving the way to the automated and efficient design of novel single-molecule magnets and coordination compounds in general.

\section*{Results}

The results section is organized as follows: i) first the fundamental ingredients of the proposed genetic algorithm will be illustrated, and ii) a first simple implementation of the method will be benchmarked against brute-force high-throughput sampling of compounds assembled from a pre-compiled list of organic ligands. Then iii) a machine learning pre-screening tool will be introduced, and finally iv) applied together with an improved encoding of ligands able to go beyond predefined lists. 

\textbf{A genetic algorithm for magnetic molecules.} By mimicking natural gene selection, a GA refines subsequent generations of genomes to search for the optimal ones according to some criteria. Each generation is obtained by crossing over the genome of parents, also accounting for random genetic mutations, and only the fittest elements of a population are selected to pass on their genetic material to subsequent generations. The fundamental ingredients of such an algorithm therefore correspond to defining i) a fitness function evaluating the quality of a certain element of the population, ii) the nature of genes and the genome, iii) a crossover strategy to generate new generations from parents, and iv) the nature of genetic mutations. 

\textit{Fitness function.} Since we are interested in selecting molecules that exhibit large magnetic anisotropy, the ideal fitness function for our purpose will simply correspond to $-D$, so that the GA will search for large and negative values of $D$ by optimizing the fitness function. As detailed in the methods section, $D$ can be accurately computed for any mononuclear Co(II) complex through ab initio multireference methods such as Complete Active Space Self Consistent Field (CASSCF), provided the coordinates of the complex are known. 

 \begin{figure*}[tp]
    \center
    \includegraphics[width=1\linewidth]{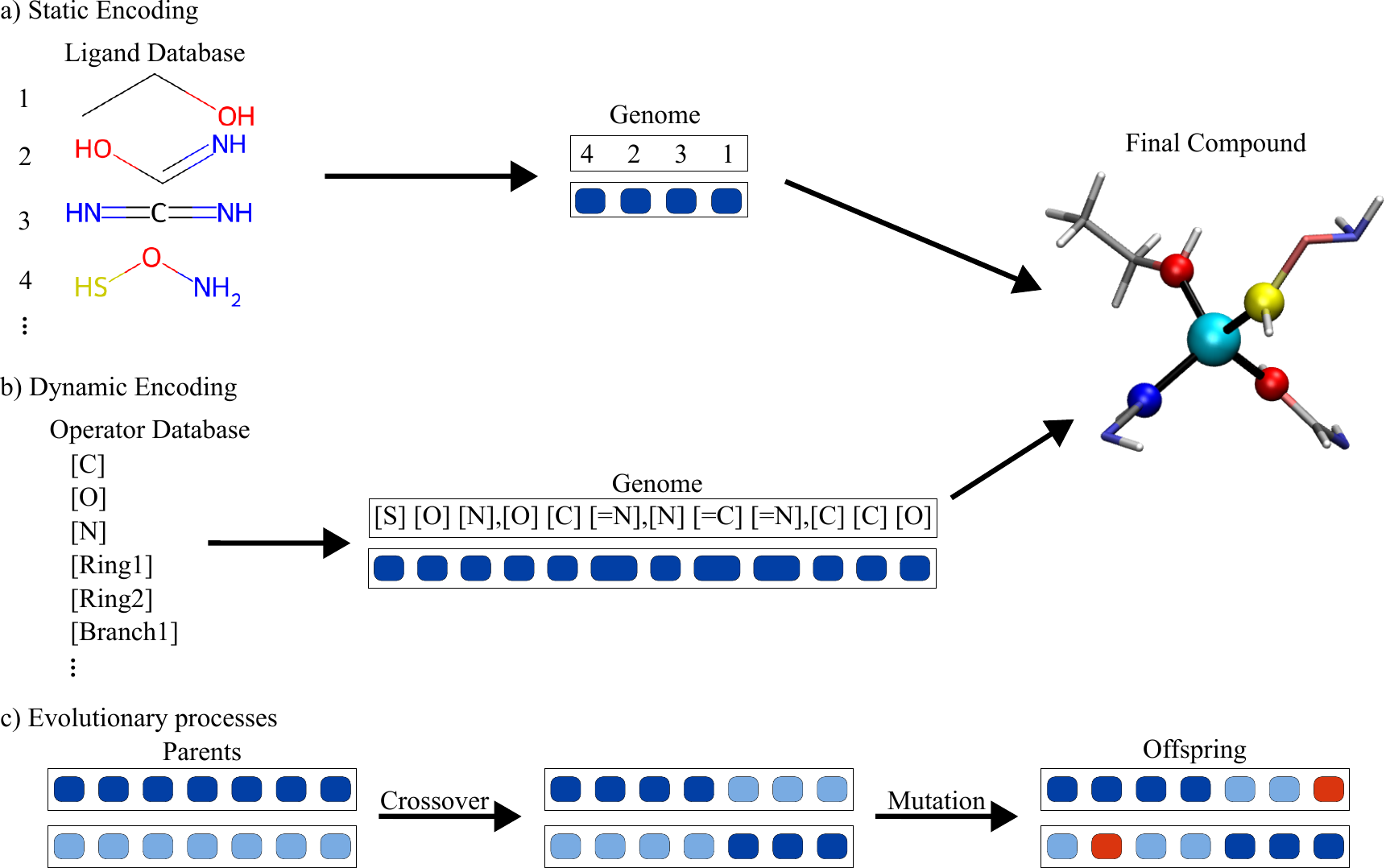}
    \caption{\textbf{Schematic representation of the genetic algorithm elements.}  (a) A prepared list of ligands is used as the basis for the algorithm and indexed (top left). The genome consists of the indexes of the individual ligands involved, with each index being one gene indicated by a blue box (top center). The final compound is then created based on the 3D-coordinates of the ligands, as seen on the right with the metal core and its first coordination shell neighbors highlighted. (b) The database for the dynamic encoding consists of individual operators, each being responsible for either adding an atom of a specific element and charge or creating rings or branches within the molecule (left). The genome is a list of these operators making up SELFIES, separated into parts for the individual ligands, and finally assembled into its 3D-representation presented in the right. (c) On the left, two parent genomes are shown, with their corresponding genes colored depending on the parent. Crossover for both encodings consists of the splitting of the genome of two parents, and cross assembly of the resulting fragments, as shown in the middle. Mutation (marked in red) consists of the replacement of a single ligand with another from the database for the static encoding, and the replacement or removal/addition of a single operator from the database for the dynamic encoding.}
    \label{GA_schematic} 
\end{figure*}

\textit{Genes and genome.} Several alternatives are available for representing a molecular complex in terms of genes. In this work, we investigate two variants, referred to as static and dynamic encodings in the following. In the static encoding approach, each ligand coordinating the Co(II) ion will correspond to a gene, and the collection of all the ligands for a given molecule will define the corresponding genome (see Figure \ref{GA_schematic}a). This encoding is implemented through a pre-compiled list of genes, e.g. a list of potential ligands is preselected and a given molecule's genome will simply correspond to the indexes of the ligands coordinating Co(II). In the dynamic encoding version, as can be seen from Figure \ref{GA_schematic}b we instead use a text string-based approach to explicitly represent organic ligands. In particular, Self-referencing embedded strings (SELFIES) are used to convert a molecular topology into a string of text \cite{KrennSelf2020}. In this approach, the SELFIES string is broken into individual operators, where each operator either adds an additional atom or creates a ring or a branch within the molecule. In this framework, each gene corresponds to one operator, while the full sequence of operators corresponds to the genome of a molecule. The usage of SELFIES ensures that any produced string of operators corresponds to a theoretically valid molecule. For simplicity, the SELFIES are generated in a way that the first token always resembles the atom that later connects to the central ion.

\textit{Crossover, mutations, and selection.} Given the population at a certain generation of the algorithm, the elements exhibiting the top 50\% values of fitness function are selected for crossover and mutation to form a  new generation, as schematically represented in Figure \ref{GA_schematic}c. Random pairs of genomes are selected for crossover among the fittest elements of the population. The crossover is achieved by splitting each genome into two parts and cross-recombining the fragments of a given pair to create a new pair of offspring molecules. In the static encoding case, the crossover then simply corresponds to mixing the list of ligands' indices of two complexes, whilst in the dynamic encoding, the crossover will involve generating a new string of text, where parts of the new string are inherited from one parent, and the rest from the other. Importantly, any SELFIES corresponds to a valid molecule by construction, greatly improving the chances of success of crossover to 100\% in the dynamic encoding case. Mutation in the static case corresponds to the replacement of one ligand with a random one from the database, whereas in the dynamic encoding a single operator is either removed, added or replaced with another from the list to allow for different string lengths. During both crossover and mutation, the first token of each parent's gene is always preserved, effectively protecting the connecting atom from changing or being lost.

\textbf{Genetic algorithm vs random sampling.} The first task we set out to do is to benchmark the use of genetic algorithms against a random sampling of ligands. To achieve this, we build on recent results from some of the present authors in using high-throughput simulations to screen four-coordinate Co(II) complexes\cite{Compass}. This dataset, named CObalt-based Magnetic Properties from Ab initio Structural Studies (COMPASS), was generated by selecting 208 chemically diverse organic ligands found in existing mononuclear Co(II) complexes, and automatically assembling 15,547 stable four-coordinate compounds with two different ligands each. Compounds were initialized in tetrahedral geometry, but geometry optimizations resulted in a wide range of coordination motifs, including distorted square planar, see-saw and octahedral. To tackle the same problem with genetic algorithms, we use the static encoding strategy, where the id of each ligand will span from 1 to 208, and the genome is composed of only 2 genes, one per type of ligand. As the genetic algorithm produces new generations, only ligands from the original list are used for mutation, so that the value of $D$ used to define the fitting function is already available from the previous high-throughput study. 

\begin{figure}[t]
        \centering
		\includegraphics[width=1.1\linewidth]{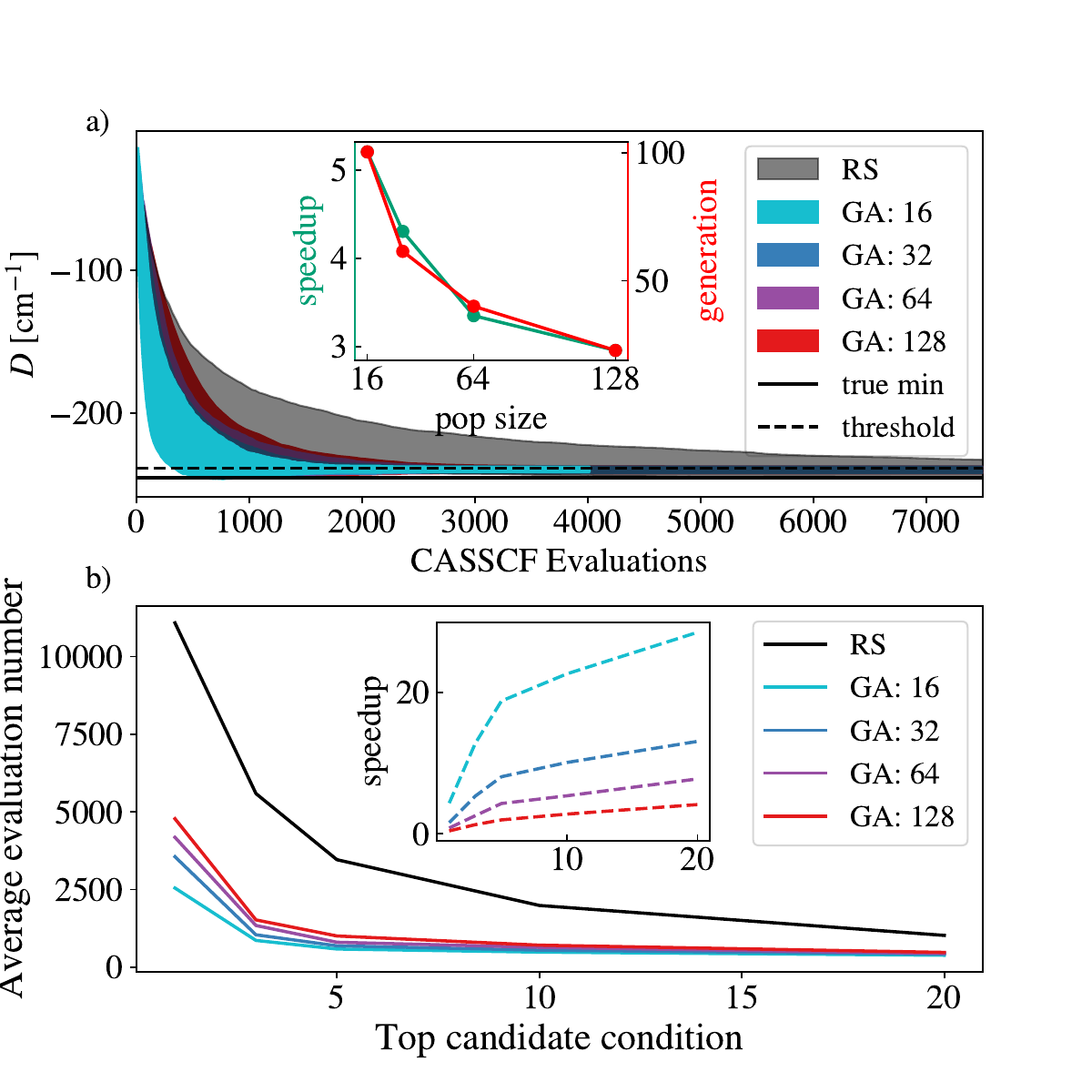}
        \caption{\textbf{Genetic algorithm performances.} (a) Comparison of minimum $D$ growth for the random sampling approach and GA with different population sizes, performed on the COMPASS test set and averaged over 1000 runs. The termination criteria is set to find one of the top three compounds in the set. The inset shows the speedup and average number of generations needed for termination of run for different population sizes. (b) Comparison of compound evaluations needed for random sampling and GA with different population sizes, versus the termination criteria—finding one of the top 1,3,5,10 or 20 compounds within the COMPASS dataset. The inset shows the speedup that GA with different population sizes enables compared to the random sampling.}
        \label{fig:benchmarking}
\end{figure}

We run multiple GA calculations, each initiated with a random selection for their initial population. Each calculation is then stopped when either the top-Nth compound in the dataset has been identified or 500 generations have been completed. For comparison, we perform a similar study in which the compounds are sampled from the full  COMPASS database in random order. Both approaches are averaged over 1000 simulated runs. The GA simulations are run for different population sizes between 16 and 128. The computational cost is then measured by the average number of different compounds that the algorithm has selected before achieving the stopping criteria. This in practice corresponds to the number of times electronic structure methods would need to be used to characterize a new molecule, adding up to the computational cost of the entire discovery process. The speedup of the algorithm is defined as
    \begin{align*}
        \frac{\text{\# of unique compounds queried by GA}}{\text{\# of compounds queried by the random sampling}}.
    \end{align*}
    
The results of these simulations are reported in Figure \ref{fig:benchmarking}, where the top panel shows the change in the optimal value of $D$ over simulation time. The average speedup of GAs over random sampling and the required number of generations needed for termination are reported in the inset of the top panel for a stopping criteria of finding one of the best three molecules in COMPASS. The advantage of GA over random sampling can clearly be seen in the average growth of anisotropy across all tested population sizes. The improved performance of the GA with smaller population sizes, compared to larger ones, can be attributed to the tendency of larger populations to retain more suboptimal candidates. This leads to less refined generations and ultimately reduces the overall quality of the gene pool. Indeed, in the extreme case of a very large population (on the order of the total search space), the algorithm would become equivalent to a random sampling. However, it should be noted that the number of generations required to achieve the convergence decreases for larger population sizes, pointing to the presence of an optimal trade-off between the number of generations and the total number of electronic structure simulations that must be executed. 
\begin{figure*}[t]
    \centering
    \includegraphics[width=.8\textwidth]{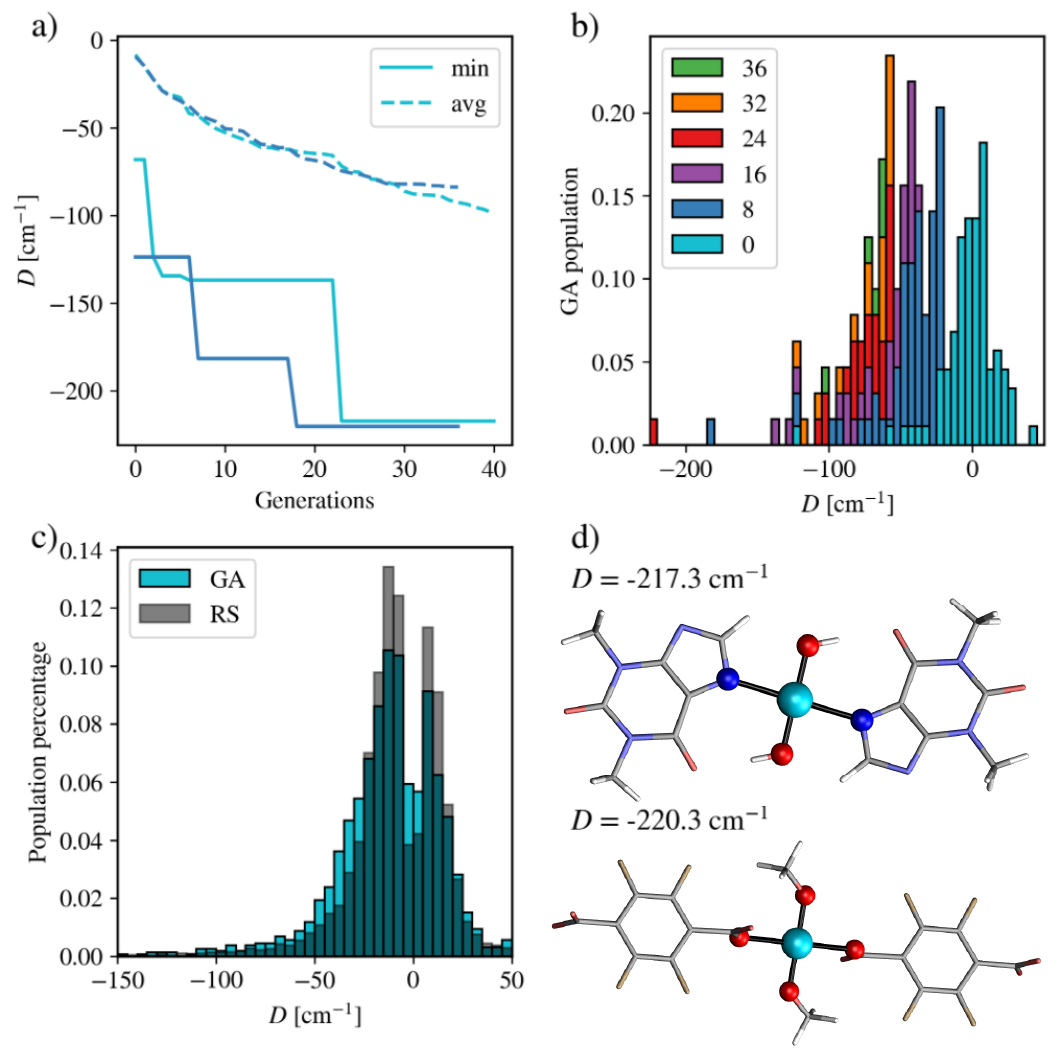}
    \caption{\textbf{Results for the GA with static encoding.} a) Minimum and average anisotropy growth of the static run. The minimum curves of both runs have a very jagged growth, as expected, while the average continuously goes down. However, both reach the < -200 cm$^{-1}$ regime. (b) Temporal evolution of GA-produced compounds and their anisotropy. The GA population at each generation roughly follows a skewed normal distribution with later generation concentrating and converging towards higher $D$ values through a steady progression. (c) Accumulative distribution of compounds to compare the GA and random sampling performances, with the GA showing a higher relative population in the medium negative regime. (d) Top performing compounds. None of the top compounds preserved the original tetrahedral geometry, however, all show reasonable optimized structures.}
    \label{fig:static}
\end{figure*}
In Figure \ref{fig:benchmarking}b we instead show the average evaluations needed for different termination criteria, with the inset showing the corresponding speedup compared to a random sampling. As it can be appreciated from these results, the advantage of GA over random sampling increases as the termination criterion is relaxed, particularly so for small population sizes, which achieve up to a factor 20 of speedup when targeting one of the top-10 compounds across the entire COMPASS dataset.

\textbf{Genetic algorithm with static encoding.} Now that we have demonstrated the advantage of using genetic algorithms over random sampling to screen the chemical space spanned by COMPASS, we attempt to tackle a much larger chemical space hardly manageable with full high-throughput screening. To do this, we select a list of 678 organic ligands from crystallographic databases using the same strategy that led to COMPASS and detailed in the Methods section. If all possible tetra-coordinated Co(II) complexes, with two different ligands each from this dataset were stable, this would generate a chemical space of 229,503 molecules. Importantly, this test, not only explores a much larger chemical space than COMPASS, but it also executes the method without a pre-compiled list of complexes. As detailed in the Methods section, we assemble and simulate with electronic structure methods a new group of molecules at each generation.

Figure \ref{fig:static}a shows the growth of the minimum and average anisotropy over GA generations. As it is common for GA, the optimal value does not necessarily improve at each generation but proceeds in big steps every once in a while. However, the average value of $D$ improves more smoothly, showing that, indeed, the overall population improves generation after generation, while breakthroughs are rare events. To better characterize how the population evolves over different generations we compute a histogram of $D$ at different generations. As it can be seen in Figure \ref{fig:static}b, as the GA run progresses, the distribution of values of $D$ across the population shifts towards negative values, signaling that the algorithm is successful in selecting genes able to optimize the figure of merit. It is also interesting to compare the evolution of the cumulative results, i.e. the distribution of $D$ over all compounds identified since the start of the GA run. This is reported in Figure \ref{fig:static}c together with the distribution coming from COMPASS, after normalizing for the total number of molecules in the two sets. The cumulative distribution shows a less marked improvement of values against random sampling as it includes early generations, where no significant optimization had already occurred. Even if the latter are ultimately discarded as non-interesting compounds, they contributed to the overall computational cost needed to achieve outstanding results at later generations. However, it can be clearly seen that the population of value towards the negative tail is larger for the GA results, highlighting the fact that the GA calculation is overall able to improve the sampling of relevant areas of the chemical space when compared to random sampling.

Finally, Figure \ref{fig:static}d reports the structure of the top-2 candidates identified by the static encoding GA. The present results are coherent with findings of the lower-scale high-throughput study that led to the COMPASS database. The 4-coordinate Co(II) compounds with the largest negative zero-field splitting $D$ are not found among the most common tetrahedral complexes, but instead exhibit distorted square planar or see-saw coordination motifs. The emergence of large magnetic anisotropy in these compounds is directly related to the partial unquenching of the orbital angular momentum $L$, which enhances spin-orbit coupling between low-lying electronic states. As previously noted in linear 2-coordinate Co(II) complexes \cite{Long2018}, large values of $L$ are due to the presence of nearly degenerate $d$-orbitals along with a significant contribution of non-Aufbau configurations in the ground-state electronic wavefunction. In the case of perfectly linear coordination, this behavior is linked to an orbital energy hierarchy such as $E(d_{z^2}) > E(d_{xz}, d_{yz}) > E(d_{x^2-y^2}, d_{xy})$, with the ground-state wavefunction predominantly composed of the non-Aufbau configuration $(d_{z^2})^1(d_{xz}, d_{yz})^3(d_{x^2-y^2}, d_{xy})^3$, resulting in an orbital angular momentum equal to 3. For the most promising candidates found in the present study, a similar effect arises due to the presence of strong $\pi$-donor ligands along the axial direction combined with weak-field ligands in the equatorial plane. The resulting ligand field leads to a 3$d$ orbital ordering $E(d_{z^2}) > E(d_{xz})\sim E_(d_{yz}) > E(d_{x^2-y^2}) \sim E_(d_{xy})$, where the ground-state wavefunction carries non-zero orbital angular momentum due to the  near-degeneracy of $d_{xz,yz}$ and $d_{x^2-y^2,xy}$ orbitals and a substantial non-Aufbau character of the electronic wavefunction. It is remarkable how effectively the genetic algorithm identifies and selects genes associated with this chemical design principle. Consequently, it tends to favor ligands coordinated through strong $\pi$-donor atoms, such as oxygen, which aligns with the trends observed in the COMPASS dataset.

\textbf{Machine learning pre-screening.} Now that we have demonstrated the ability of GA, even in the simplest static encoding formulation, to provide a computational advantage over random sampling, we are interested in pushing further the efficiency of this methodology. The first of two additional strategies in this direction concerns the use of machine learning as a computationally inexpensive method to pre-screen molecules based on being predicted to have a large negative value of $D$ or not.

 \begin{figure}[t]
    \center
    \includegraphics[width=\linewidth]{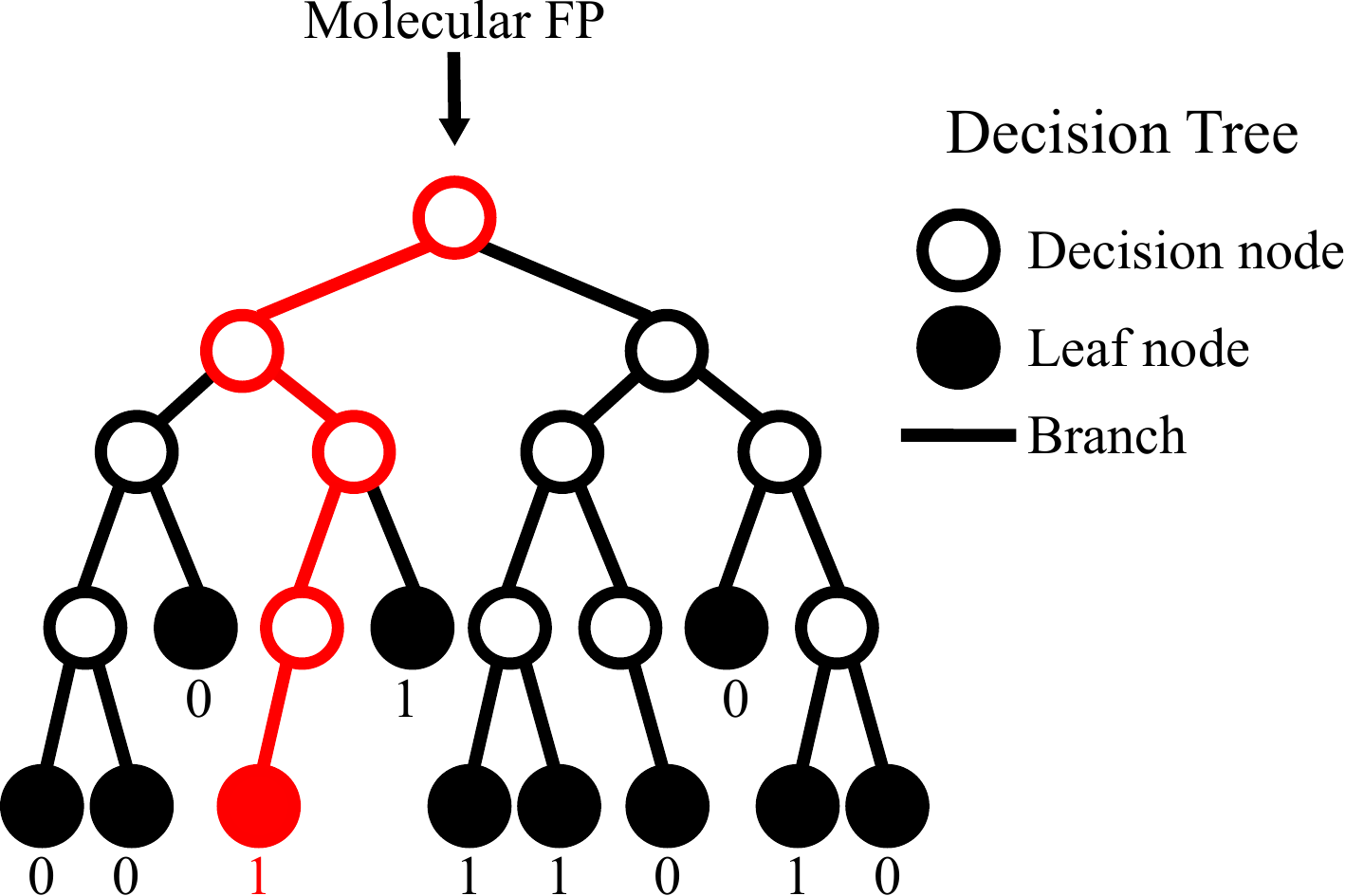}
    \caption{\textbf{Schematic depiction of an individual decision tree in a random forest.} The individual elements of a molecular fingerprint (FP) are used at each decision node to gain information about the distinction between Class 1 and Class 0. This process creates multiple branches, forming a hierarchical structure of nodes that progressively increase class purity at each split, eventually reaching the terminal nodes (leaves). The red path highlights the decision route followed by a sample molecular fingerprint as it traverses through the tree to reach a leaf node associated with Class 1.}
    \label{decision_tree} 
\end{figure}

Here, we explore the use of a Random Forest Classifier (RFC) at each GA generation to select molecules that are most likely to have a large negative $D$. Only these will be used for geometry optimization and multireference calculations to determine $D$ accurately, thus reducing computational overheads. As already used in the context of magnetic molecules by Rajaraman et al. \cite{Rajaraman2024}, RFC is a powerful machine learning technique that uses an ensemble of decision trees to classify data into desired (Class 1) and undesired (Class 0) categories. 

We start by converting string-based representations of molecules into binary vector representations, commonly known as molecular fingerprints, using the Extended Connectivity Fingerprint (ECFP) family \cite{RogersExtended2010}. Details about the fingerprints generation and the Python package used for model implementation are provided in the Methods section.
Each molecular fingerprint is then fed into a collection of decision trees, where each tree is composed of decision nodes and leaves (see Figure \ref{decision_tree}). Given a fingerprint, the decision nodes guide a path through the tree by evaluating individual elements of the fingerprint vector against learned threshold values, one at a time. Based on the outcome of each comparison, the molecule is directed down one of the branches, continuing this process until it reaches a leaf node. Each leaf represents a classification outcome, namely whether the molecule is likely to have a large $-D$ value (Class 1) or not (Class 0). The thresholds used by the decision nodes are automatically determined during the model's training by optimizing classification accuracy across the training set. Once all trees in the forest have provided a classification for a given molecule, the Random Forest Classifier aggregates the results via majority voting, assigning the final class based on the most frequent prediction among all trees.

\begin{figure}[tp]
    \center
    \includegraphics[width=\linewidth]{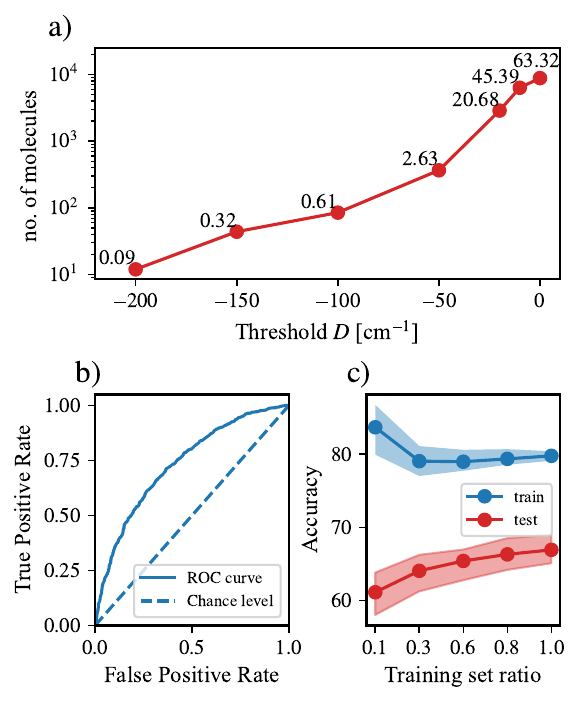}
    \caption{\textbf{Random forest pre-screening results.} (a) Number of molecules in Class 1 (i.e., molecules with magnetic anisotropy below the specified threshold) plotted against the threshold values investigated in the dataset. The percentages next to the markers within the plot indicate the proportion of Class 1 molecules relative to the total number of molecules. (b) Receiver Operating Characteristic (ROC) curve for the RFC model. The curve illustrates the model's performance compared to a random classifier. (c) Learning curve: The accuracy of the RFC model vs the training set size averaged over 100 calculations. The horizontal axis represents the proportion of the training set relative to the total COMPASS dataset. The shaded area represents the variability in accuracy across 100 calculations. }
    \label{RF_ROC} 
\end{figure}

To enable pre-screening with RFC, we define an anisotropy threshold and divide the database into two classes: desired and undesired, allowing the GA to discard structures that do not perform well. Although the ideal focus is on novel molecules with significant negative $D$ values, we note that the $D$ values in COMPASS roughly follow a normal distribution centered around zero. Less than 1\% of the total population exhibits $D$ values below -150 cm$^{-1}$, reflecting the rarity of coordination compounds with large energy barriers (see Figure \ref{RF_ROC}a). Consequently, classifying the data into categories such as below and above -150 cm$^{-1}$ results in severe class imbalance, which impedes effective ML analysis \cite{japkowicz2000imbalanced}. We achieve a more balanced classification by setting the threshold at $-10$ cm$^{-1}$, where the class ratios are 45:55 (see Figure \ref{RF_ROC}a). 
To train the model, we choose a 80\%-20\% dataset split, where the subsequent sets are used for training and tests. Considering the slight class imbalance, we identify a class weight ratio of 1:1.25 to be optimal for this data set. Further details about the remaining hyper-parameters used for training the RFC are provided in the Methods section.
Upon testing, we find that the RFC model exhibits an accuracy of 67\% and an Area Under the Curve (AUC) score of 73\% as presented in Figure \ref{RF_ROC}b. AUC indicates the effectiveness of the RFC model in distinguishing between Class 0 and Class 1. Additionally, we evaluate a learning curve to assess the relationship between training set size and model accuracy for both training and test data. As shown in the Figure \ref{RF_ROC}c, the training accuracy (blue curve) consistently remains higher than the test accuracy (red curve), indicating a degree of overfitting. However, as the training set increases in size, the model gradually improves its ability to generalize to unseen data.
This trend is further supported by the shaded regions, which represent the variance in accuracy across multiple runs. The variance is notably higher for smaller training sets, where overfitting is more pronounced and the model's performance is highly sensitive to the specific samples used in training. As the training size increases, both training and test accuracies rise, and the variance decreases—suggesting improved reliability and robustness in the model's predictions.

In addition, we test the RFC model using a set of unique compounds identified by the static GA. The new test set encompasses 678 new ligands, resulting in 2,111 novel compounds beyond the COMPASS. Measuring the performance of the RFC on this new set of molecules, we find that the model achieves an overall accuracy of 60.1\% with 57.4\% recall and 76.9\% precision. Precision measures how accurately the RFC model identifies Class 1 instances. It is calculated as the ratio of correctly predicted Class 1 cases (true positives) to the total instances classified as Class 1 (true positives + false positives). Recall, on the other hand, measures how many of the actual Class 1 instances were correctly identified. It is calculated as the ratio of true positives to the total actual positive cases in the data set (true positives + false negatives). With a high precision of 77\%, the majority of evaluations in the GA correspond to true Class 1 cases. However, the low recall results in the discarding of 43\% of Class 1 molecules. That said, given the low threshold of $-10$ cm$^{-1}$, the wide spread of $D$ values classified as 1, and the rarity of top-performing structures, the likelihood of discarding the best structures remains very low. 
In comparison, we achieve a threefold speedup, defined as the ratio of evaluation calls (predicted Class 1) and the COMPASS dataset size, and a significant gain in computational efficiency, representing an excellent trade-off.

\begin{figure}[h!]
    \centering
    \includegraphics[width=0.7\linewidth]{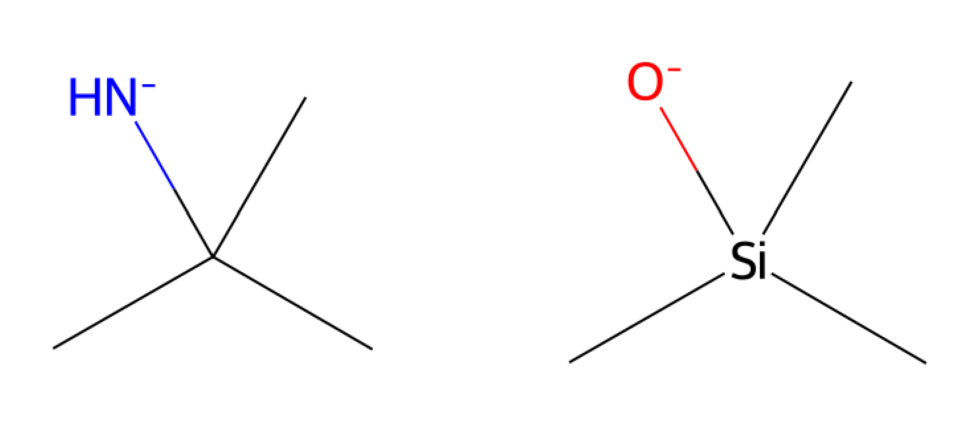}
    \caption{\textbf{Fixed ligands for dynamic GA runs.} Molecular graphs of the two static ligands selected for dynamic GA runs, chosen from the most frequently occurring ligands among the top 100 high-performing compounds present in COMPASS. }
    \label{fig:semi_ligands}
\end{figure}

\begin{figure*}[t]
    \centering
    \includegraphics[width=.8\textwidth]{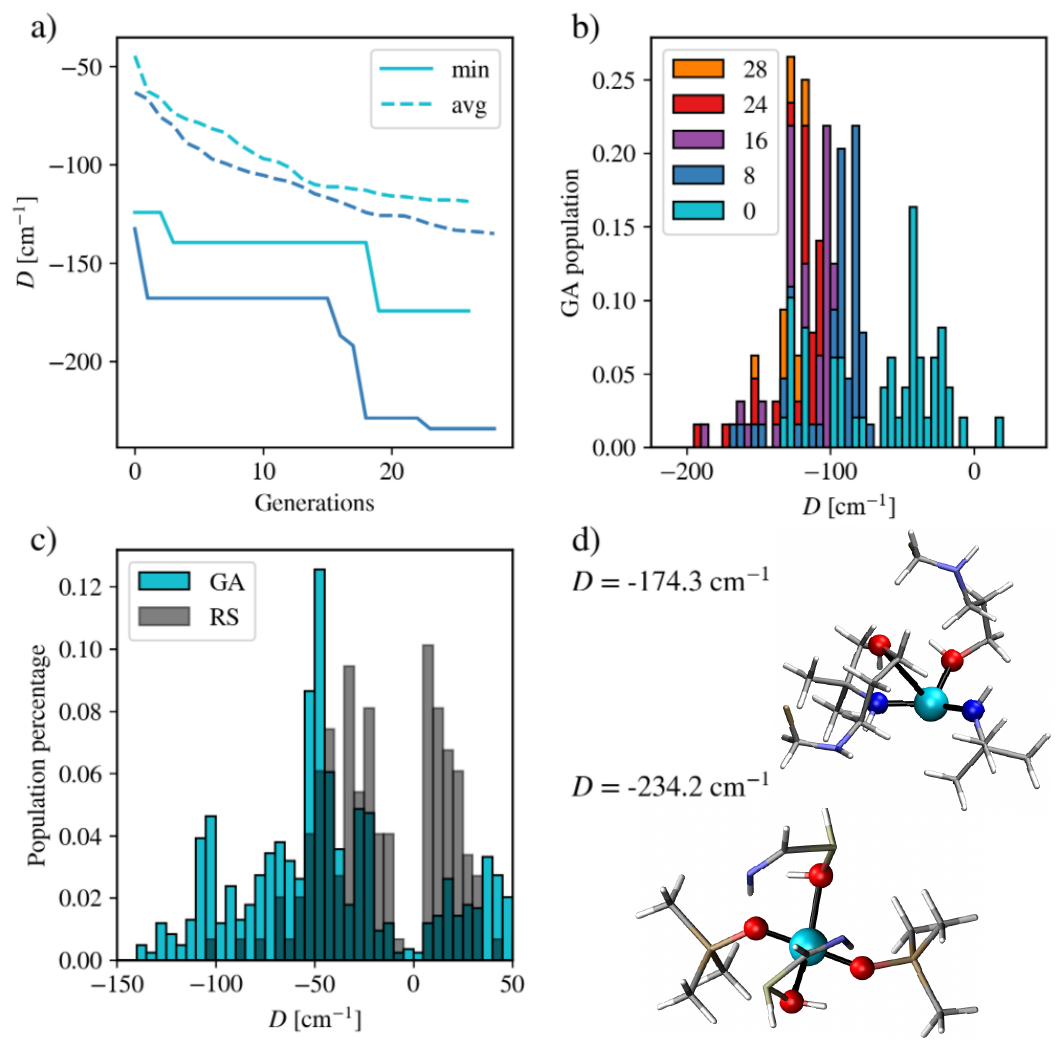}
    \caption{\textbf{Results for the GA with dynamic encoding.} (a) The dynamic GA presents a faster progress and convergence starting from an informed guess using two top-performing ligands as the fixed genes and using the RFC pre-screening. (b) The GA population is more dispersed compared to static GA with the majority of compounds reaching $<-100$ cm$^{-1}$ as early as 16th generation and a noticeable portion of the compounds accumulating in the $[-200,-100]$ cm$^{-1}$ region. (d) Top compounds, again with strong divergence from the original tetrahedral coordination. }
    \label{fig:semi}
\end{figure*}

\textbf{Genetic algorithm with dynamical encoding.} In what follows, we explore a second implementation of the GA, which, as anticipated, employs a dynamic encoding of ligands. Differently from the static encoding already discussed, genes do not correspond to the entire ligands but to fractions of them, as identified by the tokens of the strings of text used to represent them. In this implementation, ligands do not come from a pre-existing list but are generated by the algorithm in real time, leading to the exploration of a much broader, effectively infinite, chemical space. Moreover, we also exploit the RFC model to pre-screen compounds generated at each generation of the dynamic GA run and only retain those compounds that are likely to have a $D< -10$ cm$^{-1}$.

However, this formulation of the GA, although potentially much more powerful, comes with technical intricacies. In particular, we note that the value of $D$ depends on the interplay between the two ligands in the compound, one pair creating a strong crystal field in a given direction and the other pair instead minimizing it in the orthogonal direction. As such, the crossover of individual ligands results in a very unstable optimization process. To slightly ease this issue, we here explore the optimization of a single ligand, repeated twice around the Co(II) ion, while the other ligand, also repeated twice, is kept constant. As a result, the GA learns what structure combined with the static ligand maximizes the magnetic anisotropy in the compound. For this, we scanned the COMPASS dataset and selected two individual ligands from the most frequent ligands in the top 100 best-performing compounds (see Figure \ref{fig:semi_ligands}). These ligands are then used for two completely separate runs to investigate the magnetic behavior of Co(II) compounds and the impact of the new encoding on the GA performance. 

To analyze the performance of the dynamic algorithm,  we first show the behavior of the minimum and average anisotropy growth in Figure \ref{fig:semi}a. Compared to the static run, the minimum behavior is very similar, the breakthrough points are again between 15-20 generations. However, the average reaches significantly better values at a faster pace, showing the advantage of this encoding. Once it finds a good ligand, it can produce similar ligands and further optimize the compound. Figure \ref{fig:semi}b further supports this analysis, showing that the overall distribution of $D$ values moving towards very large negative ones very rapidly with the passing of generations. Next, we take a look at the distribution of the resulting compounds in Figure \ref{fig:semi}c. It is clearly visible that pre-selecting one the two ligands leads to an automatic drop of 50 cm$^{-1}$ in mean values of $D$, as shown in grey. However, the combination of this pre-selection, with the RFC model predictions, leads to an additional shift to the left in early generations, and the GA leads to significant growth in the $<-100$ cm$^{-1}$ regime, marking the advantage of this approach over brute force visible. Finally, similarly to what observed in the static run, the GA naturally identifies ligands able to stabilize a distorted square-planar or see-saw structure (see Figure \ref{fig:semi}d), which is now known to lead to an unquenched angular momentum and record-large negative $D$ values.

\section*{Discussion and Conclusions}

The design of SMMs with long relaxation time is a fascinating challenge that has kept the scientific community engaged for over 30 years. Much progress has been made in recent years in developing first-principles open quantum system methods able to provide a complete rationale for magnetic relaxation in SMMs, making it now possible to identify all fundamental ingredients supporting long relaxation times, and leaving the identification of chemical strategies for their implementation as the ultimate challenge toward the actual design of new SMMs. 

Our results prove that computational approaches are able to efficiently screen the chemical space of coordination compounds in search of magnetic molecules with record magnetic anisotropy, and therefore provide a solution to the severe roadblock imposed by resource-expensive experimental investigations. Importantly, the proposed method introduces several advantages over previous ones: the GA at the core of the approach is demonstrated to be able to find shortcuts in the chemical space, leading to an improved performance over random sampling. Once this inherent advantage of GAs is combined with a machine-learning pre-screening of candidates and dynamic ligands encoding able to access a virtually infinite chemical space, the method achieves optimal values of magnetic anisotropy in just a few generations. Such efficiency, combined with the ability to sample organic ligands outside of any pre-built database, makes this approach extremely appealing for the discovery of novel chemical motifs not easily achievable through serendipitous synthesis. 

Although our proof-of-principle study already shows excellent performance, further methodological improvement is possible and desirable. For instance, whilst the RFC helps speed up the selection of compounds generated by the GA, there is room for improvement and a machine learning model able to quantitatively predict magnetic anisotropy would represent an important step forward. Machine learning models for magnetic systems have been underexplored and will require an in-depth study of both descriptors and model architectures to become a reality. Importantly, the generation of large and diverse datasets as the ones reported here, provides the means to perform such studies. Interestingly, machine learning could lead to improved variants of GAs themselves. For instance, L. Chen and Y. Li recently proposed a scheme where machine learning is used for molecular discovery, improving the sampling of relevant areas of the chemical space \cite{chen2025uncertainty}. 

Another area of possible improvement is the consideration of multi-objective fitness functions. Here we have focused on optimizing the parameter $D$, which is known to be strongly correlated with magnetic relaxation, but that is not the only parameter determining the performance of an SMM. For instance, the rhombicity of the zero-field splitting is linked to the probability of observing quantum tunnelling of the magnetisation, which would undercut the benefits of large negative $D$ values. Similarly, the nature and energy scale of low-lying molecular vibrations are important features that must be considered to reduce the impact of spin-phonon coupling. Finally, although far from trivial, the possibility to steer the generation of molecules toward compounds that can be easily synthesized is among the most pressing and challenging aspects that need to be tackled in the future for this method to achieve the promised impact. 

Last but not least, it is important to stress that the generation of novel SMMs presented here serves as a general testbed for the generation of any coordination compound with complex magnetic and electronic properties. The method reported here can be deployed for any other class of coordination compounds and properties with minimal modification. For instance, the generation of optimized catalysts \cite{bhat2023coordination,malinowski2020application}, OLED \cite{bizzarri2018sustainable,li2022investigation}, or pharmaceuticals \cite{singh2023recent,mohammed2020medicinal}, would require a virtually identical setup, where a mononuclear coordination compounds electronic properties are tuned through a systematically improved choice of its ligands.

In conclusion, we have shown here that new SMMs with target properties can be automatically and efficiently generated through a synergistic use of genetic algorithms, multireference quantum chemistry and machine learning. This method is proven to be able to automatically lead to molecules with non-trivial coordination geometries and record magnetic anisotropy in just a few optimization steps and by scanning a virtually unlimited chemical space, ultimately demonstrating that the generation of coordination compounds with desired magnetic or electronic properties is now within reach.

\section*{Methods}

\textbf{Ligands scraping from crystallographic databases.}
The static encoding requires a list of pre-generated ligands to fill its database. This is taken directly from the COMPASS dataset (208 ligands) for benchmarking or extracted in a similar fashion from the Cambridge Structural Database (678 ligands) for the on-the-fly simulations. While this does not guarantee that it will form a stable compound with cobalt and the other ligands involved, it does indicate that the compound is, in principle, synthesizable. The process involves the extraction of all cobalt crystal structures in the database and the discarding of crystals that involve multiple overlapping conformations. Next, the crystals are split into individual molecules using the software MolForge \cite{Lunghi2022}, and only molecules containing exactly one cobalt atom are retained. By removing the cobalt atom, these can be split into individual ligands, which are then sorted by denticity, size and closed/open shell ground states. We limit our selection to monodentate, closed-shell ligands with 25 or fewer atoms, excluding any ligands with metallic or heavier than bromine connecting atoms. The resulting list is then searched by a max distance selection on the principal components of the ligands bispectrum components ($J_{max}=8$, $r_{cut}=4$) \cite{bartok2013representing}, leading to a set of 678 ligands with maximal chemical diversity within the investigated chemical space. 

\textbf{Generation of new molecular prototypes.} The first step in generating new molecular prototypes in both algorithms is sourcing the ligands. In the case of the static encoding, this is straightforward, as we simply source the already optimized ligand in its lowest energy charge state from a database. In the case of the dynamic algorithm, one of the ligands is already known, but the other is only available as a SELFIES. Using the python software package SELFIES \cite{KrennSelf2020}, it is translated to a Simplified Molecular Input Line Entry System (SMILES) \cite{WeiningerSMILES1980} string, which is a common way to describe molecules through strings of text and it is interpretable by most chemistry software. We generate an initial guess of the dynamic ligand by using RDKit \cite{rdkit}, which is then optimized using the quantum chemistry software ORCA 5 \cite{neese2020orca}, which allows for the optimization of the 3D-coordinates. The connecting atom is known from the SELFIES string, and the minimum-energy charge state is calculated assuming the ligand to be closed shell.\\
In both algorithms, the rest of the pipeline is identical: The optimized ligands are connected to the central cobalt ion to generate a coordination compound using the Python package MolSimplify \cite{molSimplify}, which orients them in a perfect tetrahedral orientation around the central ion. This initial guess is then optimized again using ORCA 5 to gain a realistic geometry and analyzed for its magnetic properties.

\textbf{Electronic structure simulations.} We use the quantum chemistry software ORCA 5 \cite{neese2020orca} to perform both DFT and state-averaged CASSCF calculations. Scalar relativistic effects are treated using the Douglas–Kroll–Hess (DKH) method, with picture-change effects included up to second order to account for DKH corrections in the spin–orbit coupling operator. The DKH-def2-TZVPP basis set is employed for all atoms, except for elements heavier than Kr, for which the SARC-DKH-TZVPP basis set is used. Energies and forces are converged to $10^{-9}$ a.u. and $10^{-6}$ a.u., respectively. The active space for the CASSCF wavefunction consists of seven electrons in the five 3d cobalt orbitals. This active space is automatically selected based on the Löwdin orbital composition from DFT calculations, evaluating the contribution of Co 3d atomic orbitals to each Molecular Orbital (MO). If any of the five highest occupied MOs contain less than 30\% Co 3d-character, they are replaced with the highest-energy occupied orbitals that exhibit more than 30\% Co 3d-character. Magnetic anisotropy is extracted from the CASSCF Hamiltonian using a mean-field spin–orbit coupling operator within the framework of Quasi-Degenerate Perturbation Theory (QDPT). The state-averaged CASSCF procedure includes 10 quartet and 40 doublet states. Compounds that fail to converge in either the geometry optimization or CASSCF calculation are assigned a default value of $D=0$ cm$^{-1}$ and automatically discarded by the genetic algorithm.

\textbf{Random forest classifier} To prepare the RFC descriptors, we first employ the RDKit and PySmiles software to generate SMILES strings for individual compounds, using molecular coordinates and charges \cite{rdkit,pySMILEs}. SMILES are string-based representations of molecules where specific characters represent different atoms, bonds, and structural features \cite{WeiningerSMILES1980}. Using the string representation, we next generate circular fingerprints from the family of ECFPs \cite{RogersExtended2010} with a radius of 5. These are then printed into binary vectors of length 4096, encoding molecular substructures up to 5 bonds away from each atom in the molecule—the radius. These fingerprints serve as features for training the RFC to predict whether a novel molecule generated by the GA framework possesses magnetic anisotropy within a desired range. We use the {\it Scikit-learn} implementation of the RFC model that accepts multiple hyper-parameters including tree depth, number of samples, features and trees \cite{scikit-learn}. After thoroughly evaluating various parameters, we identified an optimal configuration with 1,000 estimators (decision trees) and a minimum sample split of 50. This means that any leaf containing fewer than 50 samples will not undergo further splitting based on features. To prevent excessive complexity and overfitting, we regulate the growth of the tree by setting a maximum depth of 32, limiting the number of levels it can expand. 

\begin{acknowledgments}
\noindent
The authors acknowledge funding from the European Research Council (ERC) under the European Union’s Horizon 2020 research and innovation programme (grant agreement No. [948493]). Computational resources were provided by Trinity College Research IT and the Irish Centre for High-End Computing (ICHEC), and EURO-HPC. 
\end{acknowledgments}




\bibliographystyle{naturemag}
\bibliography{refs.bib}

\end{document}